\documentclass[aps,prb,reprint,superscriptaddress]{revtex4-1}
\usepackage{graphicx}

\begin{document}

\title{Spatially resolved electronic structure of an isovalent nitrogen center in GaAs}

\author{R.C.Plantenga}
\affiliation{Department of Applied Physics, Eindhoven University of Technology, P.O. box 513, 5600 MB Eindhoven, The Netherlands}
\author{V.R.Kortan}
\affiliation{Department of Physics and Astronomy, University of Iowa, Iowa City, IA 52242, USA}
\author{T. Kaizu}
\affiliation{Department of Electrical and Electronic Engineering, Graduate School of Engineering, Kobe University,1-1 Rokkodai, Nada, Kobe 657-8501, Japan}
\author{Y.Harada}
\affiliation{Department of Electrical and Electronic Engineering, Graduate School of Engineering, Kobe University,1-1 Rokkodai, Nada, Kobe 657-8501, Japan}
\author{T.Kita}
\affiliation{Department of Electrical and Electronic Engineering, Graduate School of Engineering, Kobe University,1-1 Rokkodai, Nada, Kobe 657-8501, Japan}
\author{M.E.Flatt\'e}
\affiliation{Department of Applied Physics, Eindhoven University of Technology, P.O. box 513, 5600 MB Eindhoven, The Netherlands}
\affiliation{Department of Physics and Astronomy, University of Iowa, Iowa City, IA 52242, USA}
\author{P.M.Koenraad}
\affiliation{Department of Applied Physics, Eindhoven University of Technology, P.O. box 513, 5600 MB Eindhoven, The Netherlands}

\date{\today}

\begin{abstract}
Small numbers of nitrogen dopants dramatically modify the electronic properties of GaAs, generating very large shifts in the conduction-band energies with nonlinear concentration dependence, and impurity-associated spatially-localized resonant states within the conduction band. Cross-sectional scanning tunneling microscopy provides the local electronic structure of single nitrogen dopants at the (110) GaAs surface, yielding highly anisotropic spatial shapes when the empty states are imaged. Measurements of the resonant states relative to the GaAs surface states and their spatial extent allow an unambiguous assignment of specific features to nitrogen atoms at different depths below the cleaved (110) surface. Multiband tight binding calculations around the resonance energy of nitrogen in the conduction band match the imaged features. The spatial anisotropy is attributed to the tetrahedral symmetry of the bulk lattice. Additionally, the voltage dependence of the electronic contrast for two features in the filled state imaging suggest these features could be related to a locally modified surface state. 
                         
\end{abstract}

% insert suggested PACS numbers in braces on next line
% insert suggested keywords - APS authors don't need to do this
%\keywords{}
%STM, single particle tunneling (superconductivity), 74.55.+v
%STM, in study of surface structure, 68.37.Ef

%73.22.-f	Electronic structure of nanoscale materials and related systems->73.22.Dj	Single particle states
%impurities->electronic structure, 71.55.-i ->at surfaces and interfaces, 73.20.Hb

%68.37.Ef	Scanning tunneling microscopy (including chemistry induced with STM)

\maketitle

\section{Introduction}

Small concentrations of nitrogen reduce the band gap of GaAsN up to 600 meV below that of GaAs \cite{weyers1992PLdecreasebandgapGaAsN,francoeur1998PLdecreasebandgapGaAsN, toivonen2000decreasebandgapGaAsN}, despite the band gap of GaN exceeding that of GaAs by a factor of two. Strong band bowing, common for highly mismatched alloys, has been attributed to the hybridization of the localized nitrogen states with the GaAs conduction band \cite{shan1999bandanticrossingGaInNAs1,shan1999bandanticrossingGaInNAs2} or the formation of a continuum of localized states forming an impurity band \cite{zhang2000formationimpuritybandGaAsN,zhang2001spectraofstatesNinGaAs, virkkala2013spacialftcalculationNinGaAsandGaP}.  At concentrations well below 1\%~(i.e. $<$ 10$^{19}$ cm$^{-3}$)  nitrogen induces a localized state which gives a strong narrow line in optical measurements, corresponding to an energetic feature $150-180$~meV above the conduction band edge  \cite{wolford1984resonantlevelNinGaAswithPLstrain,leroux1986PLnGaAswithNcontaminations, liu1990PLNpairinGaAs, liu1990excitonsatNinGaAs, perkins1999electroreflectanceshowslevelsNinGaAs}. Thus unlike for nitrogen in GaP, the resonance level of a single nitrogen lies in the conduction band of GaAs. Additional narrow lines have been found in the photoluminescence spectra, and assigned to other states involving the nitrogen (bound excitons, N-dopant-complexes, etc.), N-N-pairs and N-clusters, several of which are situated in the band gap \cite{zhang2001spectraofstatesNinGaAs, schwabe1985PLonNinGaAsfromVPE, schwetlick19872KPLonNinGaAsfromVPE, makimoto1995sharpPLfromNlayerinGaAs, makimoto1997NpairPLatomiclayerNinGaAs, shima1997PLoniondopedNinGaAs, makimoto1997PLandabsorptionexcitonNinGaAs, saito1997PLofNatomiclayerGaAsfromMOVPE, gruning1999PLofNbandsinGaAsN, francoeur1999excitonNclustersinGaAsN, shima1999PLspectraNinGaAsbymolecularion, kita2006growthatomicNinGaAs,kita2008magnetoPLNinGaAs,harada2014PLwithmagneticfielddeltaNlayer, harada2011magnetoPLNpairs, harada2011bound}. 

Several theoretical approaches have produced estimates of the energy levels of the single N impurity, N-N-pairs and N-clusters\cite{shan1999bandanticrossingGaInNAs1, shan1999bandanticrossingGaInNAs2, zhang2000formationimpuritybandGaAsN, zhang2001spectraofstatesNinGaAs, virkkala2013spacialftcalculationNinGaAsandGaP,jaros1979empericalpseudopotentialNinGaPAs, kleiman1979restrictedKosterSlaterNinGaAs,hjalmarson1980generalTBfordeeptrapsinsemiconductors, bellaiche1996supercellconcentratedNalloy, bellaiche1997supercellconcentratedNalloy2, bellaiche1997bandgapsfromsupercellconcentratedNalloy, bellaiche1998supercellconcentratedNalloyGaAs-GaInN-GaInAs, lindsay1999twobandTBonNinGaInAs, jones1999PLandsupercellDFTonNinInGaAsN, mattila1999planewavepseudopotentialsupercellondiluteGaAsNalloy, kent2001extensivedftonNinGaPandGaAs}, including tight-binding calculations, empirical pseudopotential calculations, band anticrossing models, and density functional theory.   
The band anti-crossing model \cite{shan1999bandanticrossingGaInNAs1, shan1999bandanticrossingGaInNAs2} is very successful in explaining the observed trend for band gap narrowing at low nitrogen content, but is a quasi-periodic theory and thus does not address the spatial structure of individual nitrogen dopants. Several supercell calculations were performed \cite{bellaiche1996supercellconcentratedNalloy, bellaiche1997supercellconcentratedNalloy2, bellaiche1997bandgapsfromsupercellconcentratedNalloy, bellaiche1998supercellconcentratedNalloyGaAs-GaInN-GaInAs,jones1999PLandsupercellDFTonNinInGaAsN, mattila1999planewavepseudopotentialsupercellondiluteGaAsNalloy,kent2001extensivedftonNinGaPandGaAs}, of which Ref.~\onlinecite{kent2001extensivedftonNinGaPandGaAs} is the most extensive, including  the energy of the single N level and N-N-pair levels, the band gap over the full nitrogen concentration range and a prediction for the spatial extent of the wave function. However, even though the supercells are almost 7~nm in linear size, significant finite size effects emerge for well hybridized states close to the conduction band or resonances within the conduction band\cite{furthmuller1992supercell}.

Here, a sample with three thin N-layers between AlGaAs marker layers grown by molecular beam epitaxy (MBE) was analyzed at 77K using voltage dependent cross section scanning tunneling microscopy (X-STM). X-STM has been applied successfully in the past to image nitrogen atoms in GaAs as atomically sized features and determine the distribution of nitrogen in various structures such as quantum wells \cite{mckay2001NinGaAsatRTinXSTM1, mckay2001NinGaAsatRTinXSTM2, duca2005InGaAs-GaAsinXSTM, ulloa2008NinGaAsQWwithXSTM}. However, until recently \cite{ishida2015NinGaAspositiveinSTM} there has been little attention payed to the electronic structure of nitrogen. At negative sample bias three types of features are seen, and extensive analysis for two of them has not yet been reported. The studies at positive sample bias reveal bowtie like features similar to the ones known for acceptor like states in GaAs. The wave function measurements are compared to tight-binding calculations which in the past have produced good predictions of single impurity wave functions in various semiconductors \cite{tang2004TBwithspinandLDOSfromGreensfunction, yakunin2004spatialstructureofMninSTM,kitchen2006MnsubstitutionSTM, jancu2008STMonMn-GaAsassymetry, ccelebi2010surfaceassymetryacceptorWF, strandbert2009TBmagneticMninGaAs, yakunin2007warpingMn-GaAswavefunctionstrain,marczinowski2007electronicstructureMn-InAs,ccelebi2008STMTBanisotropicdeepacceptorsGaPandGaAs}.
The calculations of the single nitrogen wave function in GaAs reproduce the anisotropic shape of the features seen in X-STM and show the dependence of the nitrogen wave functions on depth.

\section{Method}

The sample was grown at 550$^o$C by MBE using a 1.1x10$^{18}$ cm$^{-3}$ n$^{+}$-doped GaAs wafer as the substrate and nitrogen from a radio-frequency plasma source with ultrapure N$_2$ gas \cite{kita2006growthatomicNinGaAs,harada2014PLwithmagneticfielddeltaNlayer}. After a 400nm buffer layer of 1.0x10$^{18}$ cm$^{-3}$ n-doped GaAs, 3 nm Al$_{0.3}$Ga$_{0.7}$As marker layers and N-layers were alternately grown, starting and ending with an Al$_{0.3}$Ga$_{0.7}$As layer, with GaAs spacer layers in between of at least 35 nm. N-layers were deposited by stopping the Ga flux and opening the N-flux for 2000 s. During the nitridation the As$_{2}$ flux was kept at the same flux as during the growth (1.0x10$^{-6}$~Torr). Growth was recommenced 120~s after stopping the N-flux. After the last marker layer a 250~nm GaAs capping layer was grown. The nitridation was monitored by reflection high-energy electron diffraction.

The X-STM measurements were performed bringing the sample in ultra high vacuum (UHV, pressure typically around 5x10$^{-11}$~Torr) and cleaving the sample there, revealing a (110) plane. The sample was then cooled to 77K. STM tips were made from electrochemically etching a tungsten wire, which was then further sharpened and cleaned by argon sputtering in vacuum. Sample bias was varied per experiment, while currents were kept between 10-50 pA. Images were taken in constant current mode. Illumination of the sample was used to create charge carriers in the regions between the AlGaAs barriers.

\section{Results and Discussion}

\begin{figure*}
	\includegraphics{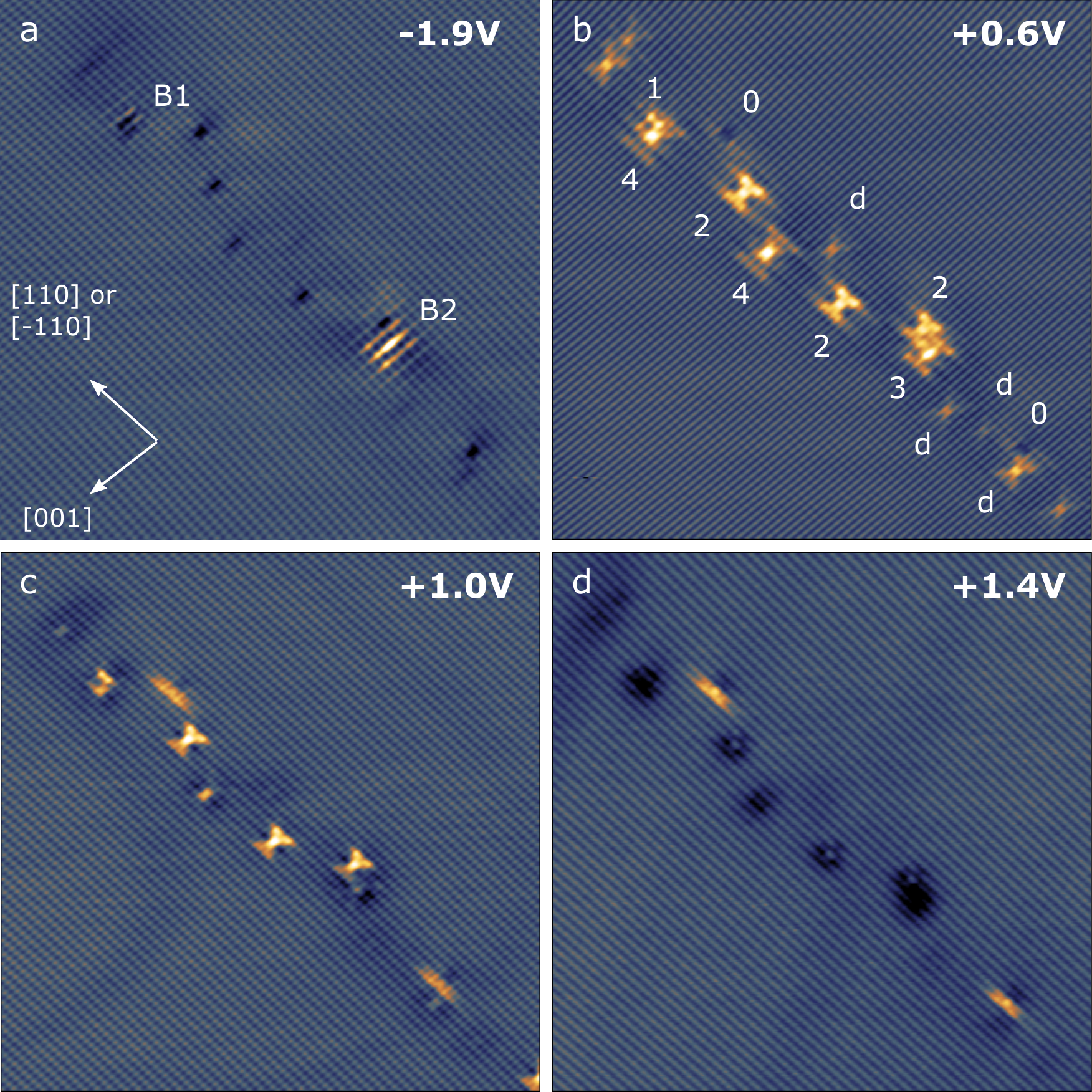} 
    	\caption{STM topography images showing a part of the nitrogen layer at (a) -1.90~V, (b) +0.60~V, (c) +1.00~V and (d) +1.40~V sample bias. The  tunnel current was 50~pA at negative sample bias and 30~pA at positive sample bias. The contrast scales are normalized to the amplitude of the atomic corrugation (giving a scale ranging over 29~pm for the lowest contrast picture and 80~pm for the highest contrast picture). B1 and B2 indicate the features containing bright contrast at -1.90~V in (a). In (b) the features are labelled `0' to `4' corresponding to the layer below the surface in which the nitrogen is located. `d' indicates features related to nitrogen situated deeper below the cleavage surface. }
    \label{voltageoverview}%
\end{figure*}

The position of the nitrogen layer is easily traceable with the help of the AlGaAs marker layers. The marker layers are not shown in the images here, since they are located 40~nm or further away. The nitrogen layer was imaged at various sample bias voltages (see Figure~\ref{voltageoverview}). At negative sample bias dark spot-like features and two types of features containing bright contrast (indicated B1 and B2) can be observed [see Figure~\ref{voltageoverview}(a)]. Dark contrast features have been observed for nitrogen in X-STM in the past with filled state imaging \cite{mckay2001NinGaAsatRTinXSTM1,mckay2001NinGaAsatRTinXSTM2,duca2005InGaAs-GaAsinXSTM,ulloa2008NinGaAsQWwithXSTM} and have been reproduced in theoretical models \cite{duan2007calculationNinGaAsSTMtopography,tilley2016calculationsNinGaAsXSTMtopography}. 
The observation of dark contrast is attributed to a depression at the surface caused by the shortened bonds between the nitrogen and its neighboring Ga atoms. These dark spots show a variation in intensity due to the variation in depth at which the nitrogen atoms are positioned with respect to the surface. The nitrogen atoms located deeper below the surface give rise to less distortion at the surface and thus a weaker dark contrast. 

The features B1 and B2 always occur around the nitrogen layer. Therefore we propose that these must be nitrogen related as well. Both features show a periodic pattern along the [110] direction, forming bar-like contrast in the [001] direction. The extent of both features differs, with the B1 feature extending about three rows in two directions and the B2 extending at least five rows in both directions.  
In previous X-STM measurements at room temperature\cite{mckay2001NinGaAsatRTinXSTM1,mckay2001NinGaAsatRTinXSTM2,duca2005InGaAs-GaAsinXSTM,ulloa2008NinGaAsQWwithXSTM} these features were not observed. A recent publication on X-STM measurements at 77~K reports a feature similar to B2, although the structure of the feature was not well-resolved\cite{ishida2015NinGaAspositiveinSTM}. 
 Discussion of both B1 and B2 will be continued later.

When imaging at +0.60~V sample bias [see Figure \ref{voltageoverview}(b)] various types of bright features can be distinguished having a complex structure with a strong anisotropy between the [110] and [001] direction. At positive sample bias these features are located at the exact positions of the dark and bright features seen at -1.90~V.  Similar features have been observed by Ishida~{\it et al.}~\cite{ishida2015NinGaAspositiveinSTM}. As will be discussed later these features are related to nitrogen substituting for arsenic atoms in different planes below the surface, where the labels 0 to 4 indicate the distance from the (110) cleavage surface in number of planes.   
Unlike the dark features at -1.90~V, all of these bright features are caused by increased tunneling current due to an enhanced local density of states (LDOS) at a resonance energy rather than the topography.

At a higher voltage of +1.00~V [see Figure \ref{voltageoverview}(c)], the 0-feature develops a strong contrast directed along the [110] direction. The structure of the other features becomes more condensed while the anisotropy in the [001] direction is preserved. 
At +1.40~V [see Figure \ref{voltageoverview}(d)] the only bright contrast is observed at the 0-feature. The other features with bright contrast at lower positive voltages now show a strong dark contrast, that unlike the localized features at negative voltage spread over multiple atomic positions. At +1.40~V the observed image deviates strongly from the observed topography at negative voltages and hence the dark contrast is attributed to an electronic origin. A reduction of LDOS to compensate for the enhancement of the LDOS at the nitrogen resonance energy is suspected. The nitrogen is an iso-electronic impurity and therefore does not introduce additional density of states when considering the integral over all energies and space. Hence a local increase of LDOS at a specific energy has to be compensated elsewhere  in space and energy. 

Comparing the measurements at positive voltages we see that the contrast intensity of the features varies with the applied sample bias. Features related to nitrogens closer to the cleaved surface have their maximum strength at higher voltages. This might be related to the states having an altered energy due to the vacuum-semiconductor-interface. More likely the tip induced band bending (TIBB) pulls up both the GaAs bands and the nitrogen resonance. The TIBB is strongest at the surface and decays away from the surface, hence features close to the surface align with the Fermi energy of the tip at higher voltages than do features farther away from the surface.

\begin{figure*}[t]
\includegraphics{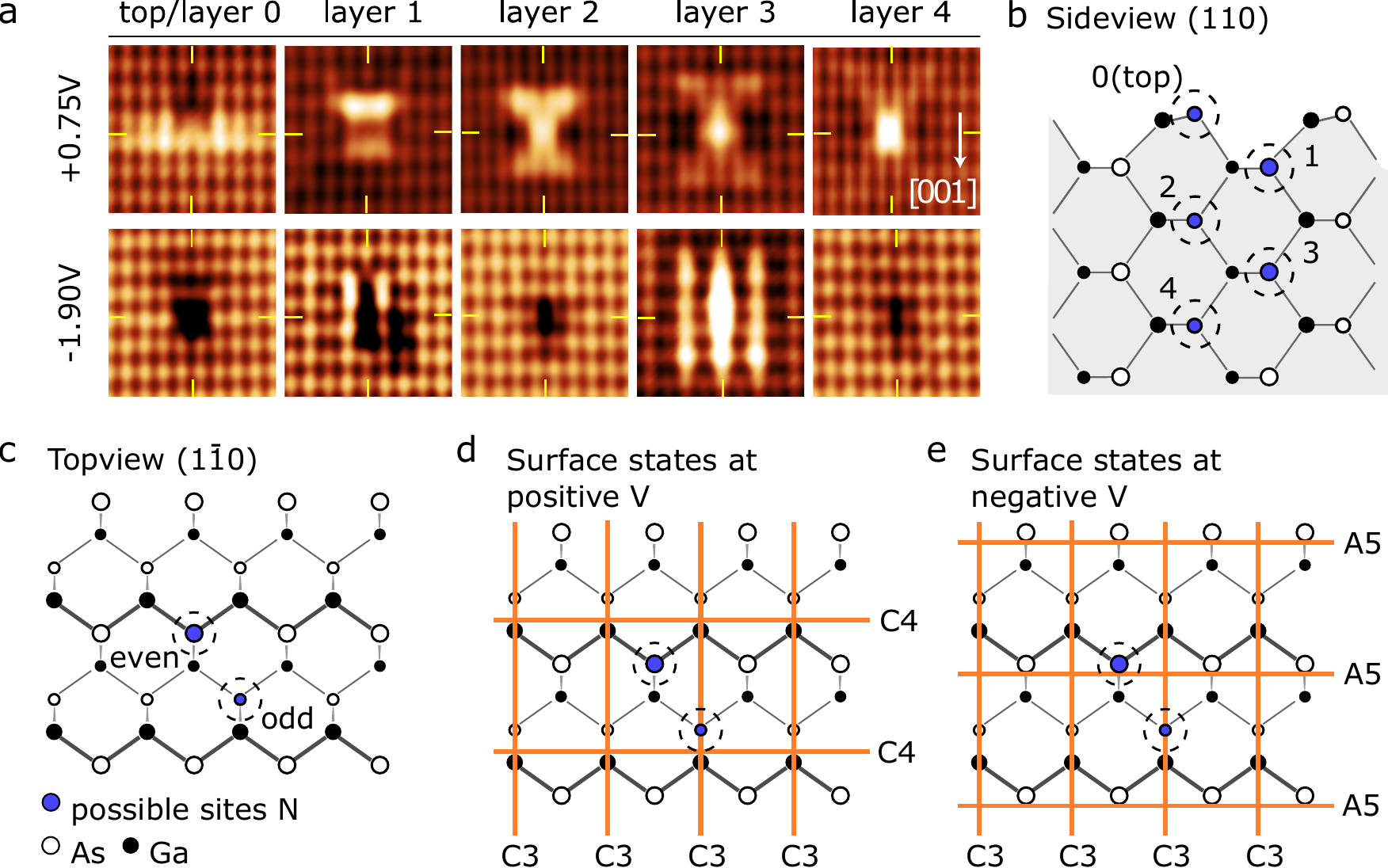}
    \caption{(a) Five nitrogen related features found at +0.75V (upper) sample bias, organized by extent in the [110] direction, and their counterparts at -1.90 V (lower). The cross-hairs indicate the position of the feature's center with respect to the surface resonances. 
 (b) Slice of the surface scanned in the X-STM along a (110) plane, showing in sideview the positions a nitrogen atom can take when substituting for the arsenic atom (c) View from the top onto the 
surface scanned in X-STM, and the two possible in plane positions for the nitrogen on substitutional sites indicated with a blue dot with dashed circle. The stronger lines and bigger atoms indicate the elevated zigzag rows in the surface. (d) and (e) show the relative position of these projections to the surface resonances imaged at negative and positive voltage, respectively.} 
   \label{layersandschematics}
\end{figure*}

In order to identify the position of the nitrogen atoms in the lattice, it is necessary to determine their position with respect to the gallium and arsenic atoms on the surface. 
As can be seen from Figure \ref{voltageoverview} the corrugation on the undoped GaAs surface depends on the applied sample bias. At -1.90 V and +1.00 V a 2D atomic grid can be observed, whereas at +0.60 V stripes directed along [110] are seen and at +1.40 V stripes directed along [001] are seen. 
The observed stripes are attributed to surface states that arise after the surface reconstruction and strongly correlate to the position of the surface atoms \cite{ebert1996surfresonancemodeSTM,engels1998surfrecandresmod}. The various surface states that were calculated were labeled A or C, according to whether the state was mainly related to the anion (A) or cation (C) sites. The maximum contribution of each of these surface states lies at a different energy, from which the numbering is derived. The spatial contribution of these states is schematically indicated in Fig.~\ref{layersandschematics}.
Around the bottom of the conduction band the C3 state is found to be dominant giving rise to stripes which are directed along [001] and centered on top of the dangling bonds of the surface gallium atoms. At low positive voltages we inject into the lower part of the conduction band via the C3 state, therefore at +0.60 V stripes along [001] are observed. At high positive voltages, like +1.40 V, we inject into the conduction band via the C4 surface resonance state, which is centered on the dangling bonds of gallium as well, but is now directed along [110]. Thus we see a voltage dependent corrugation. 
The measurement at +1.00 V shows a 2D grid for the corrugation, because at this voltage the C3 and C4 state contribute with similar weight. 

For the measurements at -1.90 V a 2D grid is observed as well. This is  remarkable because at negative voltages electrons are drawn from the valence band, which lines up with the maxima of the A4 and A5 surface states. The A4 and A5 states are centered around the arsenic surface atoms and are both directed along [110]. As was reported by de Raad~{\it et al.}~\cite{deraad2002tibbinfluencingobservedsurfacestateinSTM}, due to TIBB it is possible to observe contributions from the C3 mode also at negative voltages when tunneling close to the gap. Therefore at -1.90 V a 2D grid is formed from the C3 state and the A5 state, whose maximum is located closer to the gap than that of the A4. 

After taking into account the atomic corrugation for the clean GaAs surface the nitrogen features can be classified according to their position below the (110) surface. In the image Figure~\ref{voltageoverview}(b) and many other images at least five different contrast varieties can be observed. Figure~\ref{layersandschematics}(a) shows these features at +0.75V.  Arranging these features by the extent of the bright contrast in the [001] direction we see that their centers alternatingly fall on top or in between the imaged atomic grid [see the top row in Figure~\ref{layersandschematics}(a)]. At -1.90 V the feature centers show an alteration of position with the imaged grid as well [see the bottom row in Figure~\ref{layersandschematics}(a)]. Their positions with respect to the surface state along [001] are the same as observed for the features at +0.75V. However, the features that are centered on top of the maxima of the surface state along [110] at +0.75V fall in between the state directed along [110] at -1.90 V and vice versa. 

The nitrogen atom normally substitutes for an arsenic atom. Nitrogen atoms positioned at even numbered layers [see Figure~\ref{layersandschematics}(b)] will create a depression centered on the position of a surface arsenic atom [see Figure \ref{layersandschematics}(c)]. Hence at \hbox{-$1.90$~V} the even numbered nitrogen atoms will show up on the A5 surface resonance in the [110] direction and between the C3 resonance in the [001] direction [see Figure~\ref{layersandschematics}(d)]. Nitrogen atoms on the arsenic positions in the odd numbered layers [see Figure~\ref{layersandschematics}(b)] are not directly imaged, but will cause a distortion distributed over multiple arsenic atoms with the center of the contrast in between the surface arsenic atoms, hence in between the stripes due to the A5 state. From Figure~\ref{layersandschematics}(d) it can also be seen that these odd numbered features fall in line with the Ga atoms along [001] and therefore will be imaged on the C3 grid. 
We conclude that the states labelled with 0, 2 and 4 are indeed related to the even numbered substitutional sites and the features labeled 1 and 3 to the odd numbered substitutional sites shown in Figure~\ref{layersandschematics}(b).

The images at +0.75V have the same C3 surface state making up the rows along [001]. The center of the features hence have a similar position with respect to the [001] rows.  The C4 surface states directed along [110] are not centered on the surface gallium atoms but centered on the gallium dangling bonds.  This places the maximum integrated LDOS of the surface states next to the gallium atoms and between the zigzag rows\cite{ebert1996surfresonancemodeSTM, engels1998surfrecandresmod}[see Figure \ref{layersandschematics}(e)].  The odd numbered states will coincide more with the rows of the C4 states and the even numbered sites will fall between them producing the observed alternation in contrast with depth. 
The regularly increasing extension of the features combined with the arguments for the intensity of the dark features at negative voltage and the positioning of the features on the grid leads to the conclusion that our labelling of `0' to `4' corresponds to the ordering in distance of the nitrogen atom from the cleavage surface.

\begin{figure*}
\includegraphics[width=\textwidth]{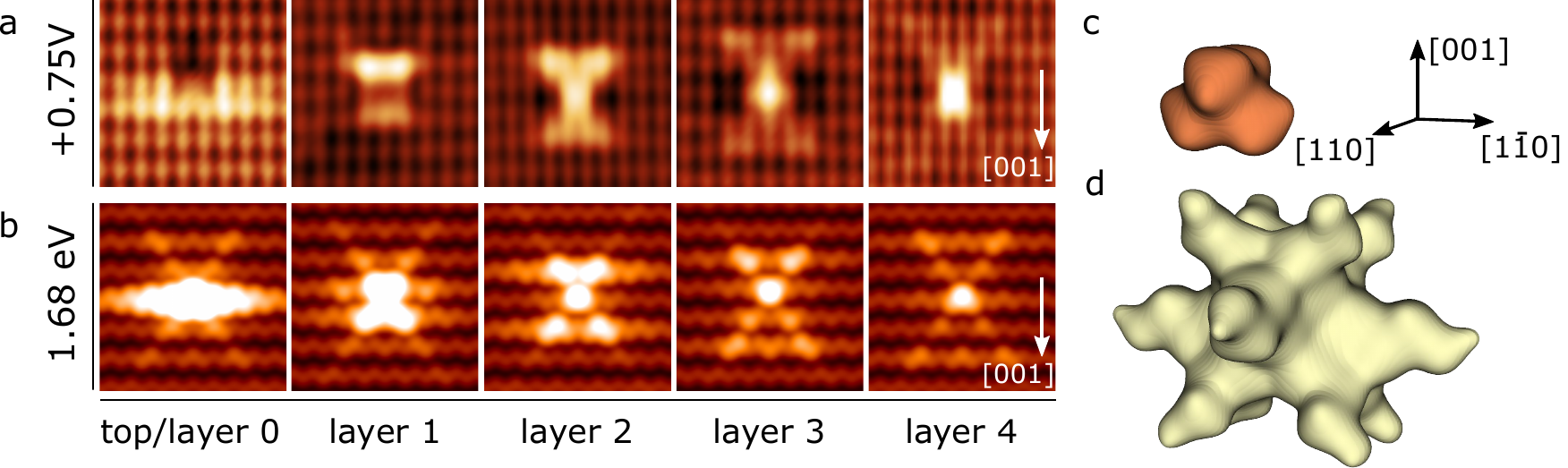}
    \caption{(a) Measured contrast at +0.75~V for the features in the 0th/top layer and below.  (b) The calculated wave functions for nitrogen atoms in the 0th/top layer and below, with an assumed STM tip width of 1.70~\AA.  The panels for calculation and measurement are about 3.6 by 3.6~nm. (c) and (d) show the calculated isodensity surfaces at 1.68 eV for the difference between the LDOS near the dopant from the background. To view sharper features in this theoretical surface the STM tip width is made smaller, 1.13~\AA.  (c) corresponds to an LDOS difference from the background of $0.12~ eV^{-1}\AA^{-3}$.  (d) corresponds to an LDOS difference from the background of $0.012~eV^{-1}\AA^{-3}$.} 
   \label{calculations}
\end{figure*}

In order to further investigate the depth dependence of the observed features, tight-binding (TB) calculations, similar to those in Ref.~\onlinecite{tang2004TBwithspinandLDOSfromGreensfunction}, were performed for a single nitrogen atom in an effectively infinite GaAs crystal.  Previous calculations of the nitrogen related states in GaAs have been performed with density functional theory \cite{kent2001extensivedftonNinGaPandGaAs, virkkala2013spacialftcalculationNinGaAsandGaP} with a reduced set of k-points and a finite supercell.  Here a sp$^3$d$^5$s$^*$ Hamiltonian \cite{Jancu1998_Empericalspds*Tight-BindingCalculationsForCubicSemiconductors} is used to describe the GaAs crystal and the Koster-Slater method \cite{Koster1954_WaveFunctionsForImpurityLevels} is used to include the effect of the nitrogen. This Green's function method is computationally efficient and does not suffer from supercell size restrictions or boundary effects.  The energy of the nitrogen resonance has been set by including an on-site atomic potential equal to the difference in s- and p-state energies between nitrogen and arsenic.  The bonds between the nitrogen and its nearest neighbors have then effectively been shortened using Harrison's d$^{-2}$ scaling law\cite{checkmail} to place the nitrogen resonance at 1.68 eV.

Figure~\ref{calculations} shows the calculated LDOS in (110) planes at various distances from the center of the nitrogen atom.  The first column is the slice through the plane containing the nitrogen atom, which would correspond with a nitrogen in the top layer of the sample surface measured in STM. The second column then shows the slice displaced one atomic plane from the nitrogen atom corresponding to the contrast 
 measured for the nitrogen in the first layer [label 1 in Figure \ref{layersandschematics}(b)] and so forth.  The calculations show a bar-like feature extending along the [110] direction for the nitrogen in the 0th/top layer and cross-like features extending in the [001] direction for cuts away from the center with an asymmetry between the two lobes. For each layer deeper into the GaAs the enhanced LDOS cross-section expands an additional row in the [001] direction.
  
Comparing the TB calculations to the measured contrast shows an excellent agreement. The series of Figure \ref{layersandschematics}(a) shows the same systematic increase of one row of bright contrast in the [110] direction as the calculations show in Figure \ref{calculations}(b) for each cut one monolayer further away from the nitrogen atom. A striking resemblance between calculation and measurement is found in the direction of extension of the nitrogen-related LDOS shape which at the surface (or 0th plane) extends in the [110] direction, while the features from other planes extend in the [001] direction.  The best correspondence between measurements and calculations is obtained with a tip width of 1.70~\AA.

As can be seen from Figure \ref{calculations}(c) and (d) the calculated isosurfaces of state density have highly anisotropic shapes. The tip width in the calculation has been chosen smaller, 1.13\AA, to make the finer features of the isosurface clearer. At a high value of the density of states the tetragonal symmetry close to the nitrogen center is recognizable. The isosurface at lower density further away shows somewhat of a preference for the $\langle$110$\rangle$ directions, similar to what was reported by Virkkala et al. \cite{virkkala2013spacialftcalculationNinGaAsandGaP}, but far less localized. The panels in Figure~\ref{calculations}(b) show the LDOS in parallel (110) planes that either cut through the N-atom, as is the case for the 0th plane, or at an integer number of atomic planes away from the N-atom. In the 0th plane the LDOS is mainly due to the two arms of the 12-fold symmetric wave function that lie in the (110) plane cutting through the N-atom. In a plane that is a few monolayers away from  the N-atom the atomic sized LDOS is mainly due to the arm that is pointing in the [110] direction. i.e. perpendicular to arms in the 0th plane. In planes at intermediate distances away from the N-atom the LDOS consists of the perpendicular [110] arm and four others arms of the 12-fold symmetric state that cut at an angle with the (110) plane.  
Strongly anisotropic shapes for the LDOS along the [001] have been reported for several acceptor impurities with levels in the band gap \cite{mahieu2005anisotropicZnBeCdinGaAswithSTM, richardella2009acceptorsinsurfaceTBandSTM} including Mn \cite{kitchen2006MnsubstitutionSTM, yakunin2004spatialstructureofMninSTM}. The anisotropy seen in acceptor states comes from the symmetry of the tetrahedral bonds in the cubic lattice and the contributing orbitals, namely the $d$-orbitals with $T_2$ symmetry and the $p$-orbitals which also have $T_2$ symmetry.  Although  nitrogen is an isoelectronic impurity, the same applies to the bonding of the nitrogen in which $p$-orbitals are contributing. Therefore, in opposition to Ref.~\onlinecite{virkkala2013spacialftcalculationNinGaAsandGaP}, we argue that the observed anisotropy of the nitrogen atom is mainly attributed to the symmetry of the surrounding potential and not the strain introduced into the lattice.

We note that for acceptor atoms that give rise to a localized state in the band gap, the spatial integral of the LDOS produces an integer value. The nitrogen atoms, however, form resonances within the continuum conduction band that locally increase the density of states around the resonance energy. The integral of this region of enhanced LDOS does not have to be an integer. The nitrogen atoms have the same number of valence states as arsenic atoms,  so  the increase in the LDOS at the resonance energy has to be compensated for  through a corresponding reduction at other energies (which are far from the X-STM measurement range).  
The local nature of the nitrogen is attributed to its small size and high electronegativity. Studies on iso-electronic substitutional impurities, like boron, which is also  electronegative compared to gallium, as well as antimony and bismuth, which cause a strong distortion of the lattice, could provide further insights on the formation of these localized isoelectronic states. A prerequisite to observe the states related to such a center, iso-electronic or not, with X-STM is that the increase in LDOS is localized enough in energy and space to significantly change the tunneling current. 

The small deviation between our calculations and measurements is due to the fact that the calculations are done for a bulk system, whereas in the experiment the nitrogen atoms are close to a semiconductor-vacuum-interface. The surface will reconstruct, deforming the layers close by and putting strain on them \cite{engels1998surfrecandresmod}. The slight additional asymmetry along the [001] direction may also be explained by surface strain, as seen for Mn acceptors\cite{ccelebi2010surfaceassymetryacceptorWF}. 
Moreover the deformation of the lattice around the nitrogen is only taken into account by changing the hopping parameters to
 the direct neighbors, without actually changing the lattice positions of the nearest neighbors or accounting for changed bond lengths to the next nearest neighbors. The nitrogen will, however, deform the lattice even beyond the first neighbors \cite{virkkala2013spacialftcalculationNinGaAsandGaP} and close to the surface this will happen in a spatially anisotropic  way \cite{tilley2016calculationsNinGaAsXSTMtopography}. These strain arguments are very likely the cause of the observed deviations and they are consistent with the observations that the features assigned to nitrogen further away from the surface are more symmetric and match the calculations better.

The calculated wave functions do not provide an explanation for the two special bright features, B1 and B2, observed at negative voltages (see Figure~\ref{voltageoverview}). At -2.50 V the bright contrast of the mixed feature B1 disappears, while the dark contrast remains unaltered (see supplementary S1). The dark contrast is in accordance to what has been reported before for the topographic contrast of a 1st layer feature \cite{ulloa2008NinGaAsQWwithXSTM,duan2007calculationNinGaAsSTMtopography,tilley2016calculationsNinGaAsXSTMtopography}. The center is positioned around a point falling in between the A4/A5 surface states and shows a dark contrast over multiple surface arsenic atoms. This strongly suggests that the observed contrast at -1.90 V is a mix of the expected topographic contrast and a bright contrast element. Feature B2 is associated with the third layer away from the cleavage surface. The expected topographic contrast for a nitrogen in this position is less strong and falls underneath the strong central bar of the bright pattern. We suggest this as the reason why at \hbox{-1.90~V} no topographic contrast element is recognizable for this feature. 

The B-features become less pronounced with more negative voltages (see supplementary S2), whereas the topography should dominate more when tunneling further from the band gap. Therefore it is likely that the bright component of the contrast stems from an electronic rather than a topographic origin.

Noticeably the bright contrast element only appears with the nitrogen atoms substituting in the odd numbered positions. Tilley et al. \cite{tilley2016calculationsNinGaAsXSTMtopography} calculated that the nitrogen atoms in those positions displace multiple arsenic atoms in the surface. 

Voltage dependent measurements on the B2 feature (see supplementary S2) suggest that the visibility of the state is related to the visibility of the C3 surface state. At less negative voltages the C3 surface state as well as the B2 feature are more pronounced. At a low positive voltage of +0.45V where the C3 mode dominates, a feature very similar to the one observed at negative voltages can be observed (see supplementary S3.2).
Therefore we propose the bright contrast of the B-features might be related to a disturbance of the C3 surface state. This would be consistent with the larger spread of the B2 feature compared to the B1 feature.  The disturbance of the surface due to the deeper lying nitrogen atom would be more extended and that of the shallower nitrogen atom would be more localized.

\section{Conclusions}
We performed X-STM measurements on individual nitrogen impurities in GaAs layers grown with MBE. The impurities were studied at different voltages. At negative voltages mainly topographic contrast appeared. At low positive voltages highly anisotropic bright shapes were observed which show voltage dependent brightness that we attribute to TIBB.  At higher positive voltages less defined dark shapes are observed, unlike the topographic contrast observed at negative voltages. This we relate to a decrease of LDOS compensating the increase of LDOS at lower energy. 
Using the  difference in extent of the observed features at low positive voltages and the atomic corrugation coming from the voltage dependent surface states, the features can be assigned to nitrogen at different planes below the cleavage surface.  TB calculations give similar anisotropic enhanced LDOS at (110) cuts through and next to the nitrogen center. These results show that the anisotropic shape of the LDOS is caused by the tetrahedral symmetry of the nitrogen atom substituted for arsenic in the GaAs lattice. Minor deviations between the experimental and theoretical contrast can be attributed to the deformation of the lattice, caused by  surface relaxation and the small nitrogen atom, which is only partially accounted for in these bulk calculations.  At negative voltages not only topographic features are observed, but for two features B1 and B2, a resonant electronic component with an alteration in the [110] direction is found as well.  These features can be attributed to nitrogen in the first and third layer away from the cleavage surface. Earlier calculations and measurements show that these are the positions in which the disturbance of the surface is the most delocalized.  Measurements at varying negative voltages show that a relation with the C3 resonant state is likely.

\begin{acknowledgments}
This work is supported by NanoNextNL, a micro and nanotechnology consortium of the Government of the Netherlands and 130 partners. This work is supported in part by an AFOSR MURI. 
\end{acknowledgments}

\bibliographystyle{apsrev}
\bibliography{bibliography-NinGaAs2-aps}

\pagebreak

\end{document}

\title{Spatially resolved electronic structure of an isovalent nitrogen center in GaAs- Supplement}

\author{R.C.Plantenga}
%\email[]{Your e-mail address}
%\homepage[]{Your web page}
%\thanks{}
%\altaffiliation{}
\affiliation{Department of Applied Physics, Eindhoven University of Technology, P.O. box 513, 5600 MB Eindhoven, The Netherlands}
\author{V.R.Kortan}
\affiliation{Department of Physics and Astronomy, University of Iowa, Iowa City, IA 52242, USA}
\author{T. Kaizu}
\affiliation{Department of Electrical and Electronic Engineering, Graduate School of Engineering, Kobe University,1-1 Rokkodai, Nada, Kobe 657-8501, Japan}
\author{Y.Harada}
\affiliation{Department of Electrical and Electronic Engineering, Graduate School of Engineering, Kobe University,1-1 Rokkodai, Nada, Kobe 657-8501, Japan}
\author{T.Kita}
\affiliation{Department of Electrical and Electronic Engineering, Graduate School of Engineering, Kobe University,1-1 Rokkodai, Nada, Kobe 657-8501, Japan}
\author{M.E.Flatt\'e}
\affiliation{Department of Applied Physics, Eindhoven University of Technology, P.O. box 513, 5600 MB Eindhoven, The Netherlands}
\affiliation{Department of Physics and Astronomy, University of Iowa, Iowa City, IA 52242, USA}
\author{P.M.Koenraad}
\affiliation{Department of Applied Physics, Eindhoven University of Technology, P.O. box 513, 5600 MB Eindhoven, The Netherlands}

\begin{abstract}
The supplement contains three figures. The first two figures show negative voltage series around bright feature B1 and B2 which hint towards a relationship with the surface states imaged in STM. While Figure 1 images both features for three negative voltage values, Figure 2 shows bright feature B2 over an extended voltage series. Both show a clear change in the imaged surface states and the appearance of the features with that. Figure 1 additionaly shows that the contrast for the B1 feature is made up out of a topographic and an electronic contribution. 

Figure 3 shows an extension of Figure 1 from the paper, adding images at +0.45~V and +1.70~V and using a different contrast. A remarkable observation is that at +0.45~V features with a strong resemblance to the features at -1.90~V can be seen at the position of the B1 and B2 features. As there is a strong contribution of the C3 surface state at +0.45~V, this observation supports the idea that the observation of the bright features is related to the observation of the C3 surface state, rather than the polarity of the voltage. Furthermore it forms a further support for the observation that the voltage at which each feature has its strongest bright contrast is different for each variety. This is especially clear for the `0' feature which shows no bright contrast at +0.45~V, but shows a brighter contrast than any of the other features at +1.40~V and +1.70~V. Thirdly at +1.70~V the strongly localized dark contrast from the topography becomes visible, giving indications that the dark contrast at lower voltages is not related to topography and hence stems from electronic contributions.                       
\end{abstract}

\maketitle

\begin{figure*}
\includegraphics{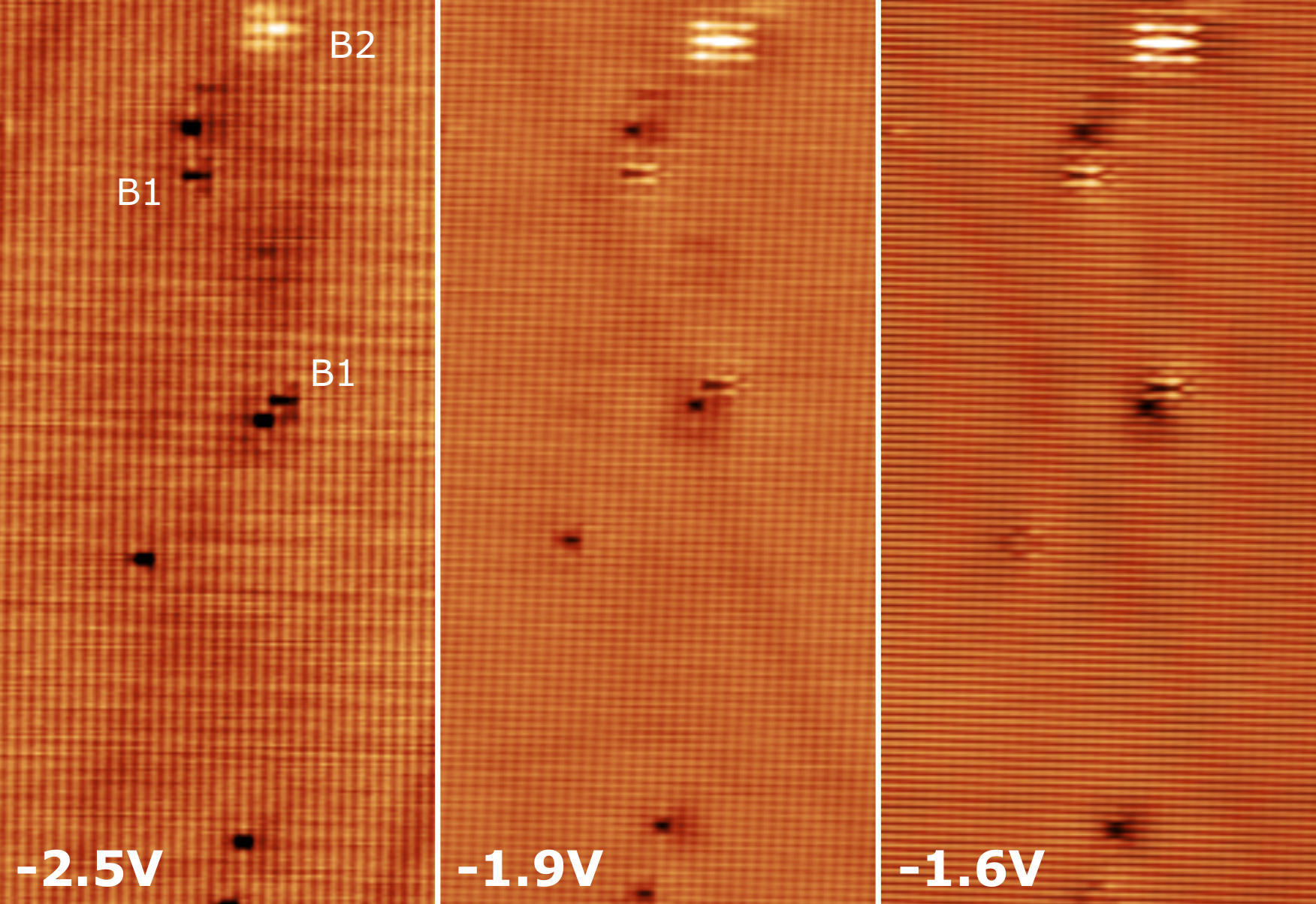}
    \caption{Part of the nitrogen layer, containing features B1 and B2, imaged at -2.50~V, -1.90~V and -1.60~V. The current is kept at 30 pA (range kept constant)} 
\end{figure*}

\begin{turnpage}
\begin{figure*}
\includegraphics{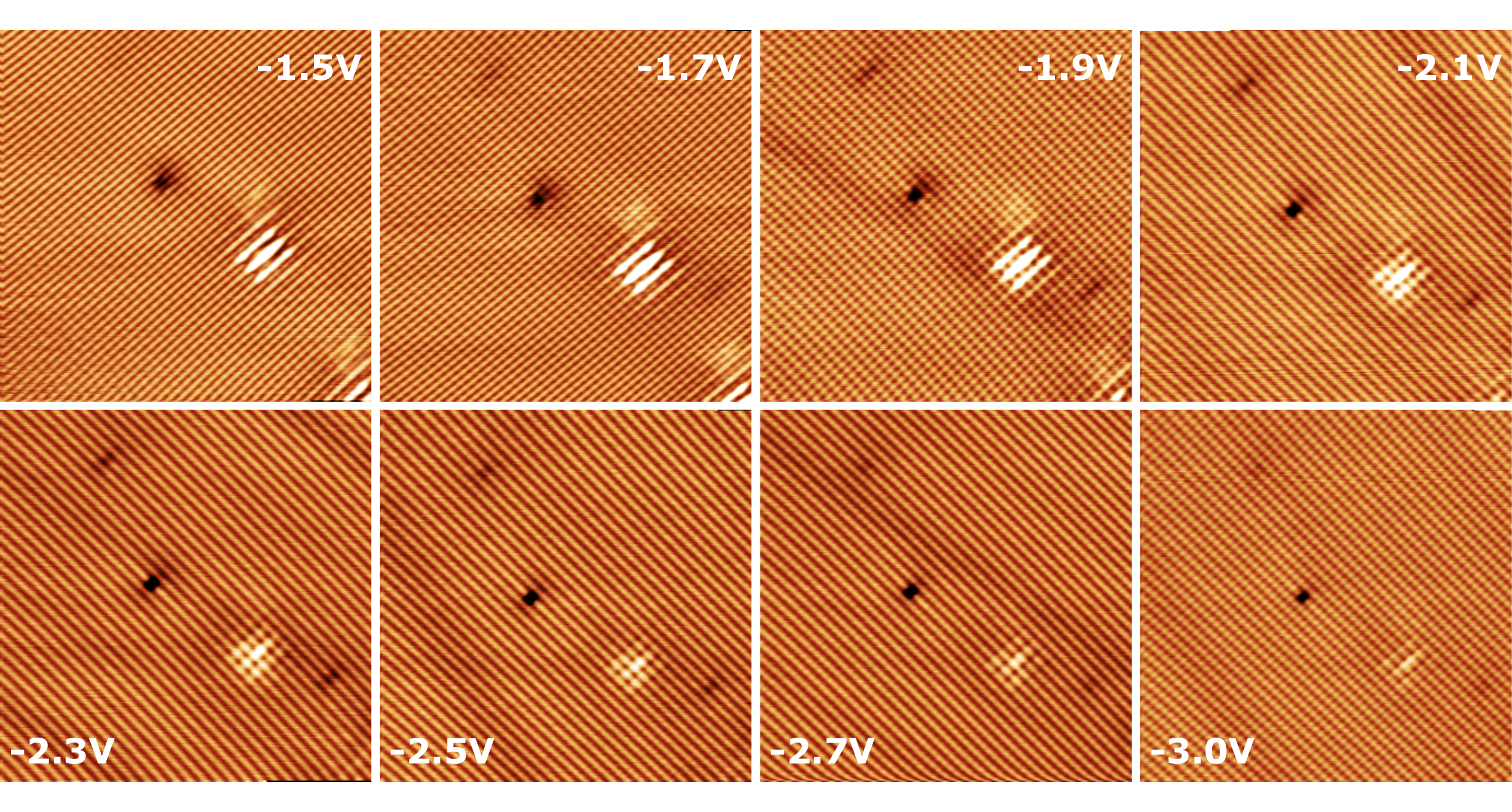}
    \caption{Voltage series of feature B2 ranging from -1.5~V to -3.0~V at currents of 30~pA. Next to a clear change in the appearance of feature B2, the alteration in addressed surface states is eminent.} 
\end{figure*}
\end{turnpage}

\begin{turnpage}
\begin{figure*}
\includegraphics{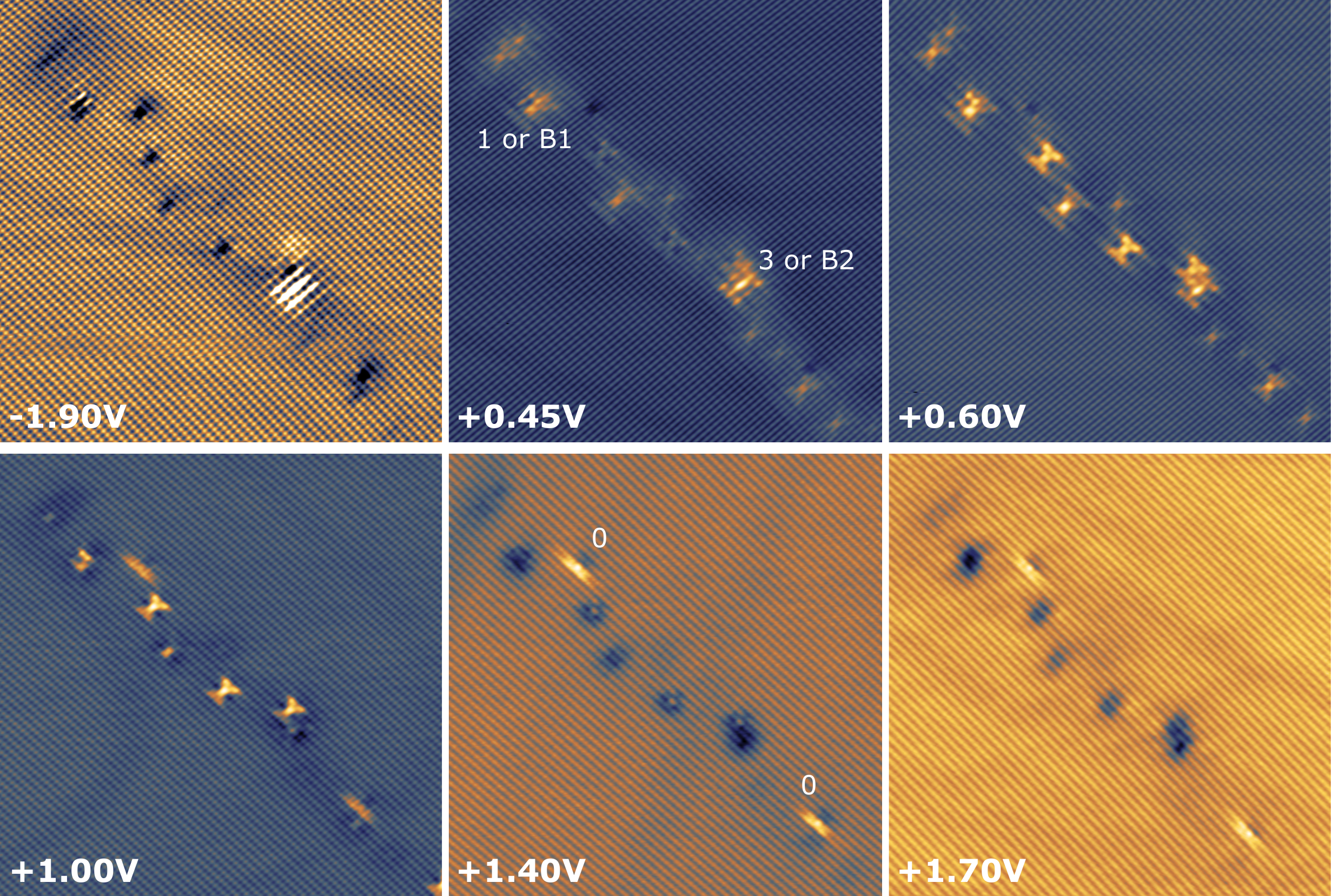}
    \caption{Extended voltage series of the series presented in Figure 1. Height range is adapted per picture for optimal contrast. At +0.45~V the features associated with the first and third layer, show a contrast resembling features B1 and B2 at negative voltages. At +1.40~V and +1.70~V the feature in the top layer solely shows an anisotropic bright contrast. At +1.70~V the measured contrast resembles the measured image at -1.90~V, with mainly topographic contributions, more than at +1.40~V.} 
\end{figure*}
\end{turnpage}